# Initiation of waves in BZ encapsulated vesicles using light -----towards design of computing architectures

Ben de Lacy Costello[1,2], Ishrat Jahan[1,2], Matt Ahearn[1], Julian Holley[2], Larry Bull[2] and Andrew Adamatzky[2]

1 Institute of biosensing technology, UWE, Bristol, BS161QY

2 Unconventional Computing Group, UWE, Bristol, BS161QY

E-mail: Ben.DeLacyCostello@uwe.ac.uk

**Abstract**

A gas free analogue of the Belousov-Zhabotinsky reaction catalysed by ferroin and encapsulated in phospholipid stabilised vesicles is reported. A reaction mixture which exhibits spontaneous oscillation and excitation transfer between vesicles was formulated. By adjusting the reagent concentrations a quiescent state with fewer spontaneous oscillations was achieved. Using relatively low power laser sources of specific wavelengths (green 532nm and blue 405nm) it was shown that waves could be reproducibly initiated within the BZ vesicles. Furthermore, despite the reduced excitability of the system overall the initiated waves exhibited vesicle to vesicle transfer. It was possible to manipulate single vesicles and design simple circuits based on a 2D validation of collision based circuits. Therefore, we conclude that this BZ system exhibits promise for computing applications based on 3D networks of vesicles.

1. Introduction

Both theoretical and experimental implementations of the BZ reaction have been widely used for computing, for example see overview in [Adamatzky *et al*. 2005]. Initial experiments centred on constructing logic gates in geometrically constrained architectures [Toth and Showalter 1995, Toth *et al*. 1994, Steinbock et al. 1996, Agladze et al. 1996 ]. The use of the light sensitive analogue of the BZ reaction further enhanced these approaches [Motoike and Yoshikawa 1999, Agaldze *et al.* 2000, Gorecki *et al.* 2003] although experimental implementations were relatively few [Gorecka and Gorecki 2003, Ichino *et al.* 2003]. Most of the approaches use the interaction of excitation waves to code the inputs and outputs of logical operations.

Adamatzky developed the ideas of collision based computing when applied to BZ computing [Adamatzky 2003 and 2004]. This approach aims to minimise architecture and logical functions are constructed via collision of excitation wave fragments in a weakly excitable system. A number of papers validated these ideas in theoretical constructs [Adamatzky 2004, Adamatzky and de Lacy Costello 2007] and experimental systems [de Lacy Costello and Adamatzky 2005, Toth *et al.* 2009] with [Toth et al. 2009] and without [De Lacy Costello and Adamatzky 2005] external control via the application of light. This idea was extended in attempting to construct a 1 bit half adder in theoretical models and in a light sensitive BZ analogue [de Lacy Costello *et al.* 2011]. It was found that in such schemes a lack of wave stability especially in experimental implementations limited the scope of this approach. Wave instabilities can however be limited by reducing functional longevity. Encapsulating the reaction in 2D disc shapes of appropriate size and connection weights enhanced the stability and allowed complex computational schemes to be derived [Adamatzky *et al*. 2011, 2011b]. Subsequently it was shown in simulations [Holley *et al*. 2011] and experiment [Holley *et al*. 2011b] that this approach allowed the construction of a wide range of logical gates including NAND and XOR. It was also shown that this modular approach could be utilised to construct arithmetic circuits in experimental systems with the first successful implementation of a 1 bit half adder [Holley *et al.* 2011b]. The great utility of this approach was that disc size, connection angle and connection weight (pore size) could all be used in tandem to control wave stability and computational outcome. The work described in this paper aims to show at least in principle that these ideas of collision based computing can be extended to the 3D case by using macroscopic vesicles containing the BZ solution.

The BZ-AOT system constitutes a water in oil reverse microemulsion with nanometer sized droplets of water incorporating the BZ reagents, stabilised by the surfactant sodium bis(2-ethylhexyl)sulfosuccinate (AOT) dispersed in octane. A number of interesting macroscopic properties have been observed based on the microscopic cooperative interactions of the BZ droplets. These include the formation of Turing Structures[Vanag and Epstein 2001], packet waves[Vanag and Epstein 2002], trigger waves[Vanag and Epstein 2001], dash waves[Vanag and Epstein 2003] and anti-spirals[Vanag and Epstein 2001b]. The main reason for these observed phenomena are the alteration of the diffusion rates of the primary activator and inhibitor species dispersed across the oil and water phases. More recently a macroscopic (millimeter sized) implementation has been reported [Symanski *et al.* 2011] which involved the encapsulation of the BZ reagents in phospholipid based vesicles. They utilised a conventional ferroin catalysed BZ system and were able to show wave transfer. They also reported the computational potential of these systems. The work presented in this paper involves changing the BZ system from a malonic acid based to a cyclohexadione (CHD) based gas free reaction [Kurin-Csorgei *et al.* 1996]. This has certain perceived advantages when encapsulating the system in vesicles. It has also been reported that the ferroin catalysed CHD-BZ reaction exhibits higher susceptibility to light than the ruthenium based analogue [Kurin-Csorgei *et al.* 1997]. This is in stark contrast to the Malonic acid based system. The paper further reports new results on the reproducible initiation of excitation waves in the ferroin catalysed CHD-BZ reaction using laser sources of different wavelengths. This is important as the control of wave initiation and transfer is a key aspect of any computational implementation.

2. **Experimental**

**2.1 Chemicals**

Sodium bromate, potassium bromide, sulphuric acid, 0.025M stock solution of ferroin and decane were purchased from Aldrich (Aldrich, UK) and used as received unless stated otherwise. Lecithin granules were used to make a stock solution of phospholipid by dissolving 1g in 100ml of decane (stock was refrigerated after use). The 0.3M stock solution of 1,4-cyclohexadione was prepared by dissolving 1.68g of 1,4-cyclohexadione in 50ml of 1M sulphuric acid at $50^{o}C$.

**2.2 Preparation of CHD-BZ encapsulated vesicles exhibiting spontaneous oscillation and vesicle to vesicle transfer**

A total volume of circa. 2.5mls of BZ solution was prepared with the following stock solutions. The system consisted of 0.1ml of 3M $H_2SO_4$, 0.235ml deionised water, 0.95ml of 1.5M $NaBrO_3$, 0.875ml of 0.3M CHD, 0.15ml of 1M KBr and 0.17ml of 0.025 ferroin. Therefore, the final concentrations were as follows: Sulphuric acid 0.47M (including conc from CHD stock), Sodium bromate 0.57M, potassium bromide 0.06M, CHD 0.11M and ferroin $1.7\times10^{-3}$.

In a 50ml beaker the distilled water, sulphuric acid and sodium bromate were added followed by the cyclohexadione solution. The beaker was then transferred to a fume cupboard and stirred using a magnetic follower. Next the potassium bromide solution was added and the solution was allowed to clear prior to adding the ferroin catalyst solution. The solution was left to stand, note that the induction period for the CHD catalysed BZ reaction at these reagent concentrations was approximately 2 hours. Once the BZ solution started to oscillate approximately 0.5mls was pipetted into a glass petri dish (4cm diameter) which contained a shallow layer of the lipid solution (<5mm in depth). The solution was agitated mechanically in order to form dispersed vesicles. If required vesicles could then be transferred to a separate reaction vessel also containing lipid solution or decane in order to undertake controlled experiments on a limited subset of vesicles.

**2.3 Preparation of CHD-BZ encapsulated vesicles for light initiation experiments**

The excitability of the system was reduced by altering the concentration of sodium bromate in the reaction mixture. The altered reaction mixture consisted of 0.1ml of 3M $H_2SO_4$, 0.335ml of deionised water, 0.85ml of 1.5M $NaBrO_3$ followed by 0.875ml of 0.3M CDM, 0.15ml 1M of KBr and 0.17ml of ferroin. Therefore, the final concentrations were as follows: Sulphuric acid 0.47M, sodium bromate 0.51M, potassium bromide 0.06M, CHD 0.11M and ferroin $1.7\times10^{-3}$.

The vesicles were prepared using the same method described above. Following formation and segregation of the vesicles the reduced vesicles were interrogated with laser light of different wavelengths (Red 650nm < 1 mW, Blue 405±10nm < 1mW and Green 532± 10nm < 1mW). Reduced vesicles from the same batch of BZ mixture were selected (based on uniform size of 1-3mm), isolated and exposed to each of the laser sources independently to assess whether localised wave initiation was reproducibly initiated. The experiments were repeated a number of times with different batches of the same BZ solution. In separate experiments vesicles were placed in close proximity within glass vessels to assess whether after initiation waves were able to transfer between vesicles.

The wave activity within the vesicles was observed using a SCopetekDCMC510 (USB 2.0)camera (5 M pixels, CMOS CHIP, brunel microscopes) attached to a microscope (Prior (James Swift). The results were recorded as still images and video sequences using SCopePhoto software.

3. Results and Discussion

3.1 The CHD-BZ-Vesicle system

The organic substrate malonic acid used by [Symanski *et al.* 2011] was replaced by 1,4-Cyclohexanedione (CHD). The original purpose for using CHD in contrast to the classical BZ reaction that employs malonic acid, is that the CHD-BZ reaction does not generate significant concentrations of gaseous products such as $CO/CO_2$. The advantage of this reaction is the ability to place it in closed systems for the lifetime of the reaction without compromising the dynamics by nucleating gas bubbles [Kurin-Csorgei *et al.* 1996]. Therefore, it seems an ideal analogue of the BZ reaction for use in experiments of vesicle encapsulation. Indeed we found that despite a relatively long induction period (compared to that of malonic acid) of approximately 2 hours (for the reported concentrations), the general behaviour of the system (spontaneous oscillations, vesicle to vesicle wave transfer etc.) within vesicles was as reported by [Symanski *et al.* 2011], although the reagent concentrations used are distinct.

Figure 1. shows the natural behaviour within a network of polydisperse vesicles. In Figure 1a most vesicles are in the reduced state. It can be seen that spontaneous oscillation has occurred in a limited number of vesicles which are in the oxidised state. In Figure 1a wave transfer between an oxidised vesicle and a reduced vesicle in the top central portion of the image can be observed. In subsequent images it can be seen how the wave is transferred from North to South through the network of vesicles. The directionality of the transfer shows that excitation is passed from vesicle to vesicle and is not a result of spontaneous oscillation. The term oscillation is used because there are no discrete wavefronts observed within the vesicles. However, because the vesicles are analogous to unstirred reactors then localisation of the excitation is apparent. This localisation means that transfer at the boundaries can be more easily observed and attributed to inter-vesicle transfer and not spontaneous initiation. Figure 1 also shows that despite the polydisperse nature of the vesicles that spontaneous initiation and wave transfer occurs in a range of sizes spanning 0.5mm-5mm. In experiments with the malonic acid system we found that small vesicles below 1mm were not subject to spontaneous oscillation. It is also obvious from Figure 1 that the path of the wave transfer is mediated by the proximity of vesicles and that this is size independent. Where waves do not transfer to vesicles in close proximity we can assume they are in a less excitable state or still in a refractory state. It can be seen in Figure 1 that vesicles that were initially oxidised start to revert to a refractory/reduced state—indicated by the colour change. However, we found that vesicles never fully regained the vibrant orange colour of the reduced state after initial oxidation. Again we can presume this is because the vesicle represents a very small volume unstirred closed reactor. Figure 1. Shows that the vesicles generally maintain good stability over the lifetime of the experiments. However, there is still some evidence of vesicle amalgamation, with the large oxidised vesicle in the centre right of the image growing at the expense of smaller vesicles. We found this to be a common problem particularly for reaction mixtures with higher levels of acidity than the ones reported (see figure 2a-f). An interesting observation was that amalgamation of vesicles was uncommon between two reduced vesicles but increasingly common where vesicles were in the oxidised state. It was also found that small vesicles (<1mm diameter) were stable with respect to larger vesicles. Indeed figure 2e shows that small vesicles were able to persist for long periods even at the interface between two large vesicles. We can postulate two distinct reasons for this observation one chemical and one mechanical. It seems likely that the chemical environment at the interfacial boundary between the vesicles will be altered during the course of the oxidation. This change in the chemical environment may temporarily disrupt the lipid membrane causing amalgamation. Secondly it is well documented that oxidation waves are capable of exerting physical force [Kitahata *et al.* 2002]. They reported the deformation of a single BZ droplet suspended in oil during excitation wave dynamics [Kitahata *et al.*]. More recently they have shown that the motion of the droplet can be controlled by using a photosensitive catalyst and a light gradient [Kitawaki *et al.* 2012].Therefore, we can postulate that vesicles already in close proximity are more susceptible to amalgamation in this system due to oxidation induced deformation and increased interfacial tension. As smaller vesicles are less susceptible to spontaneous oscillation and have more stable lipid membranes due to a favourable size/surface area ratio this explains why they are resistant to amalgamation generally. Even though amalgamation was considered undesireable for the purposes of our experiments and why we devised recipes that exhibited minimal spontaneous vesicle amalgamation, it may still be useful in computing schemes. This would be particularly true if a method of selectively controlling this mechanism were elucidated. This type of mass transfer mechanism coupled with the excitation wave transfer mechanism could be used to devise collision based computing schemes. For example if vesicles containing different reaction mixtures, activators or inhibitors of the reaction can be collided and undergo mass transfer it would provide a fine level of control over the reaction evolution. According to the paper of Kitawaki *et al.* it may be possible to use the wave dynamics coupled to an external light field to manipulate the movement of the vesicles and therefore undertake controlled collisions and mass transfer. It may be possible to exert fine control over the proximity of vesicles thus switching between an excitation wave transfer and mass transfer mechanisms.

Figure 3. shows the reconstruction of a 2 input OR gate in vesicles as simulated in 2D models of the Oregonator [Holley et al. 2011a] and implemented in 2D light sensitive experimental BZ systems [Holley et al. 2011b]. Figure 2e shows that waves transfer from both inputs into the central output cell causing oxidation of the output vesicle. Although this scheme is not fully designed it does show that the functionality of this gate should be realised in 3D vesicle ensembles. If only one of the inputs had been activated then the wave should have travelled across the ouput vesicle and subsequently transferred into the second input vesicle, thus outputting 1. If excitation waves in vesicles behave in a similar way to discrete waves in 2D experimental systems then the construction of an AND gate should be trivial. This would simply involve the addition of an output cell orthogonal to the current output. Therefore, wave transfer to this output would only occur due to the collision of two wavefronts in the current output. It is possible that if the vesicles in proximity to the gate in figure 3 had not been oxidised already that additional functionality could have been observed in the current scheme.

Figure 4a. shows the first step to constructing memory cells in 2D discs in a simulated BZ medium. In this scheme initiation of any disc will cause a wave to travel clockwise and counter clockwise resulting eventually in annihilation due to collision. This is equivalent to a memory RESET function. It is possible that if additional output cells are placed adjacent to selected cells in this memory cell then the convergence of the wavefronts would cause transfer of excitation to the additional outputs. These can then be used as inputs into additional memory cells or logical circuits. Figure 4b shows that this simulated scheme was able to be implemented in 2D light sensitive BZ media. Figure 4c shows that the same scheme can be implemented in 3D using a connected network of vesicles. An oxidised vesicle in the top right hand corner of the image initiates a wave in an input of the memory cell. In successive images this wave can be seen to transfer bi-directionally to adjacent vesicles. Eventually the two initiated waves converge at one "output" vesicle and collide and annihilate. As the adjacent vesicle to the ouput is already in the oxidised state then no transfer externally from the memory cell occurs. Once the memory cell has regained its excitability additional inputs would be possible. This idea can be expanded by the addition of diode junctions in simulation (see Holley *et al.* 2011), however the diode functionality is not proven for 3D vesicle implementations as yet.

### 3.2 Activation of the CHD-BZ-Vesicle system with visible light

This scheme uses a reaction mixture which has a reduced concentration of bromate in order to reduce spontaneous oscillations. This mixture when used for the formation of vesicles maintains equilibrium in the reduced state for a number of minutes prior to the onset of spontaneous oscillations. Therefore, it is ideal for testing the activation of waves via visible light.

Figure 5 shows the application of a red laser to a vesicle in the reduced state. There was no evidence of activation even after repeated 10 second exposures to the laser light. This finding is in contrast to previously published data [Toth *et al.* 2002] where it was found that the ferroin catalysed BZ system immobilised on a polysulphone membrane could be initiated by a HeNe red light laser of wavelength 632.8nm and power 20mW. However, it should be noted that the power of the laser source is much higher in the case of their experiments.

Figure 6 shows the application of blue/UV laser light at a wavelength of 405nm to a BZ vesicle in the reduced state. Even after 5 seconds exposure time a wave was reproducibly initiated in this system. It is clear that the initiation of the oxidation front is localised at the approximate point of application of the laser. Previous work found that wave initiation was often at the interfacial region between high and low light intensity rather than at the point of maxmimal light intensity [Toth *et al.* 2002]. Figure 7. Shows that despite using a mixture of relatively low excitability to minimise natural oscillations that an oxidation wavefront initated via the application of blue laser light is able to transfer to an adjacent reduced vesicle. This is important when considering the design of computing circuits with networks of vesicles. It is important to have a controllable system in terms of stable behaviour and wave initiation but not at the expense of inter-vesicle "communication".

Figure 8 shows that localised oxidation waves can also be initiated using a low power green laser of wavelength 532nm. It is not certain whether the initiation is caused directly by light (and is therefore dependent on the wavelength as suggested by our experiments) or local heating (and therefore dependent on the power of the laser, which although relatively well matched the rating is not specific and therefore, the exact incident power cannot be easily ascertained). It is well known that the light sensitive BZ reaction exhibits a wavelength susceptibility (also to blue light) which is linked to the activation state of the ruthenium catalyst. However, there has been limited work on the light sensitivity of the ferroin catalysed reaction [Toth *et al.* 2002]. For the purposes of this study, where reproducible initiation is required then it is not necessary to understand the exact mechanism of initiation. However, it would be an interesting future study as light sensitivity is not a well understood phenomena in the CHD analogue of the BZ reaction. Although there have been previous reports of the susceptibility of the ferroin catalysed BZ-CHD system to light [Kurin-Csorgei *et al.* 1997]

The effects of the exposure of red, blue and green laser light to an already oxidised vesicle was also investigated. Experiments were repeated and exposure times were varied but there were no visible changes to the oxidised vesicle on exposure to the red or green laser. Exposing an oxidised vesicle to the blue/UV laser, resulted in the exposed area of the vesicle being temporarily reduced, see figure 9. Upon removing the laser light the reduced area was re-oxidised . This is presumably because it remains in the vicinity of oxidised product as the application of the laser beam and thus the reversion to the reduced state is very localised. Presumably if the experiment was repeated with smaller vesicles then it should be possible to switch the state of an oxidised vesicle to the reduced state. It should also be possible to selectively disrupt wave transfer between vesicles as at this point where there is a minimal amount of oxidation at the boundary region. This is interesting from a point of view of implementing computing schemes because as much controllability as possible would be required.

Reversing oxidation in certain vesicles would be useful for resetting circuits etc. Selective online annihilation of oxidation waves would also be beneficial for implementing complex circuits.

## 4. Conclusion

This work has shown that it is possible to incorporate the CHD analogue of the BZ reaction in macroscopic mm sized vesicles. Furthermore it has shown that spontaneous oscillation resulting in vesicle to vesicle transfer of oxidation fronts is reproducibly observed in this system. It was possible to show in principal that computing schemes developed in 2D models [Holley *et al.* 2011a] and experimental systems [Holley *et al.* 2011b] reported previously could be translated into 3D systems incorporating vesicles.

Most significantly it was shown that localised oxidation fronts could be initiated by the application of laser light (wavelengths 405 blue and 532 green) within reduced vesicles which had lower natural excitability (reduced bromate concentration). Furthermore, it was shown that after initiation vesicle to vesicle wave transfer could still be observed. It was also found that application of the blue laser to already oxidised vesicles caused localised reversion to the reduced state.

The findings in this paper facilitate the aim of implementing computational schemes [such as those developed in Adamatzky *et al.* 2011a,b, Holley *et al.* 2011 a,b and Bull *et al.* 2013] in 3D networks of vesicles.  Our collaborators (http://neu-n.eu/) are currently assessing the design of BZ-vesicle based computing circuits using microfluidic devices [King *et al.* 2012].

**Acknowledgements**

The work is part of the European project 248992 funded under 7th FWP (Seventh Framework Programme) FET Proactive 3: Bio-Chemistry-Based Information Technology CHEM-IT (ICT-2009.8.3). The authors wish to acknowledge the support of the EPSRC grant number EP/E016839/1 for support of Ishrat Jahan. We would like to thank the project coordinator Peter Dittrich and project partners Jerzy Gorecki and Klaus-Peter Zauner for their initial work in this area and useful discussions.

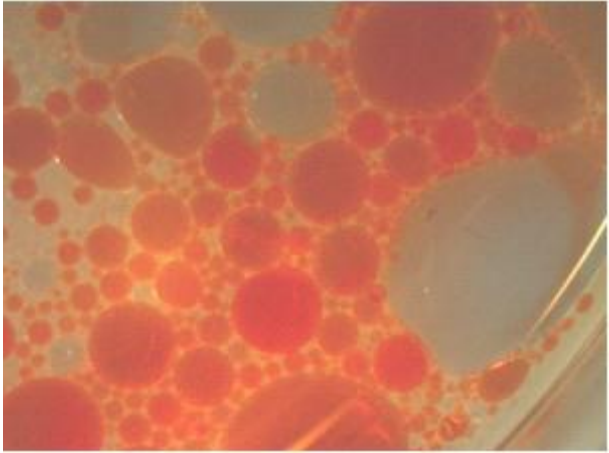

Figure 1a

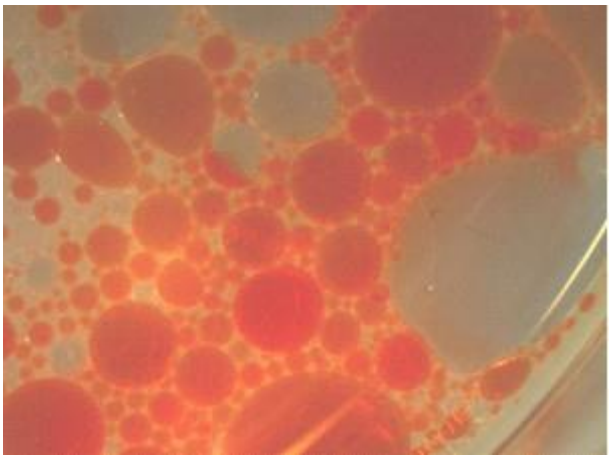

Figure 1b

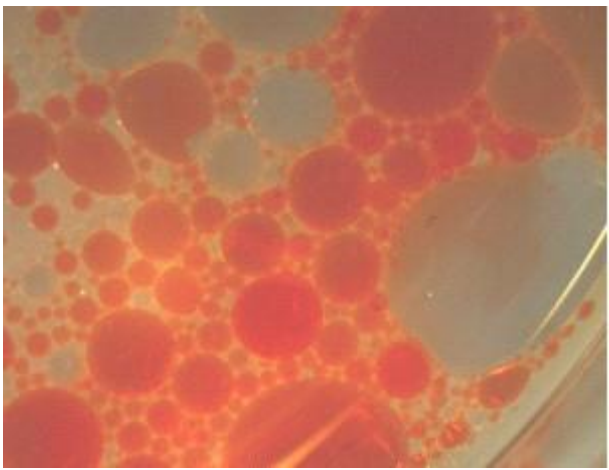

Figure 1c

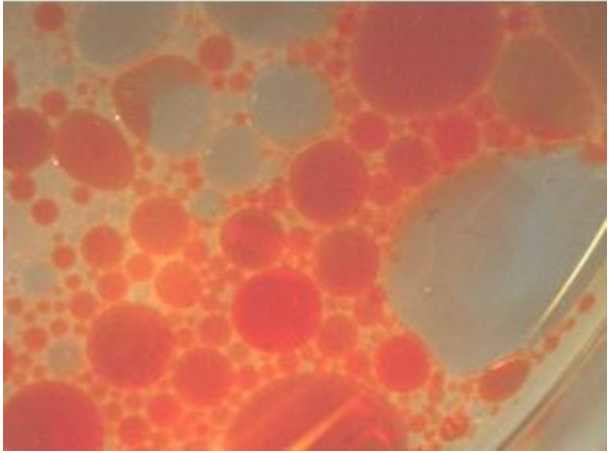

Figure 1d

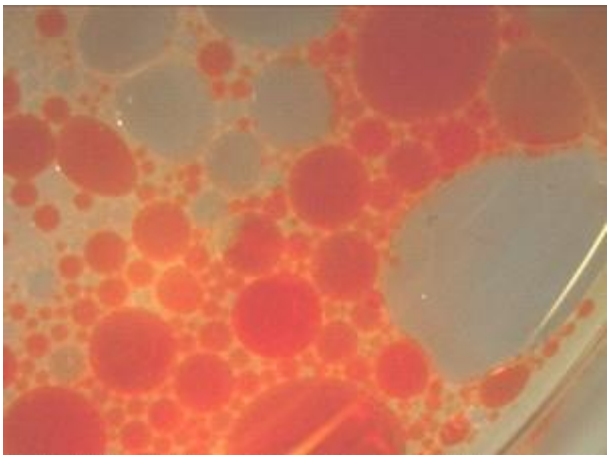

Figure 1e

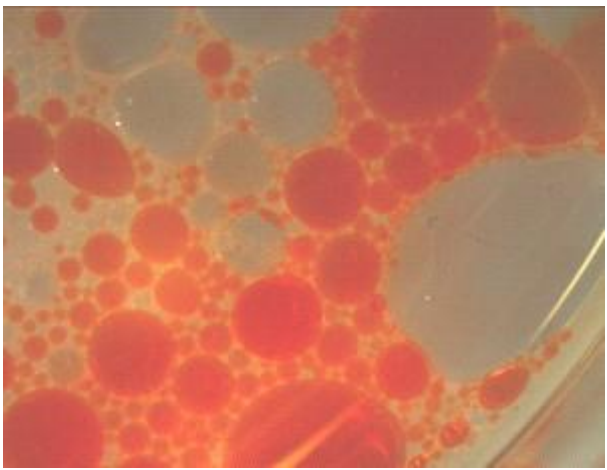

Figure 1f

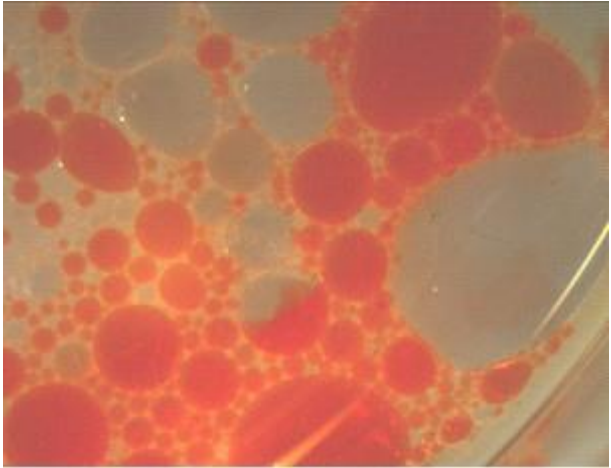

Figure 1g

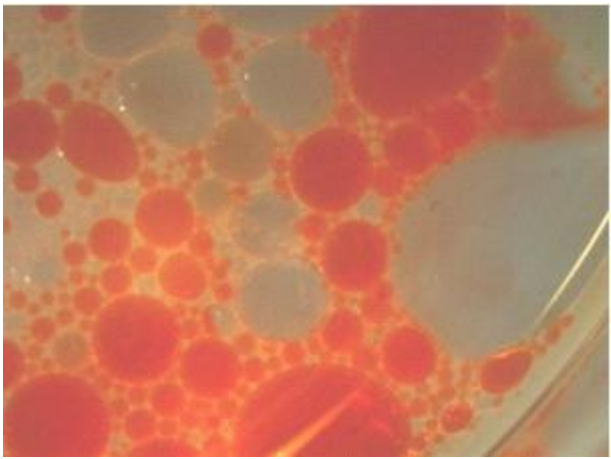

Figure 1h

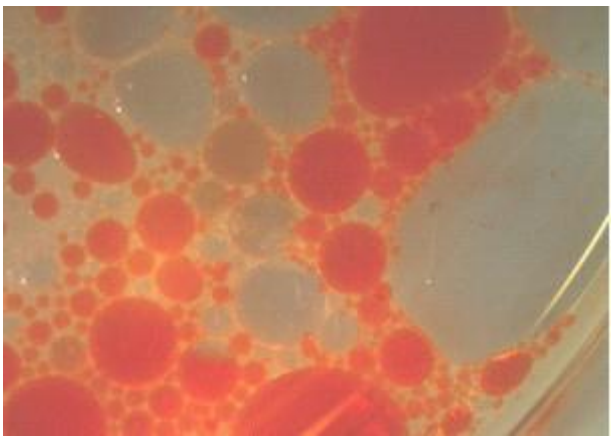

Figure 1i

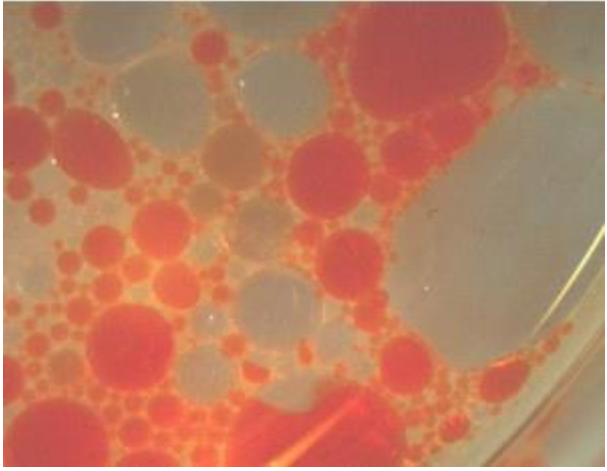

Figure 1j

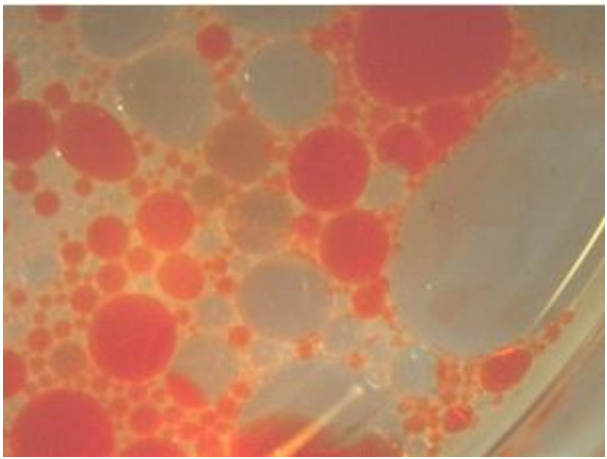

Figure 1k

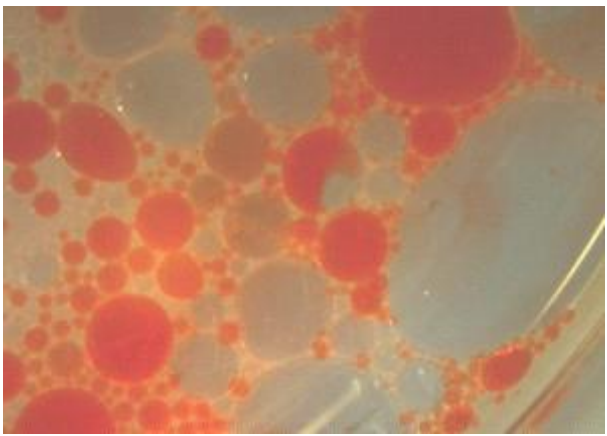

Figure 1l

Figure 1a-l. Showing spontaneous initiation and vesicle to vesicle wave transfer in a dispersed network of vesicles. A wave is travelling in the general direction north to south through images a-l. Images were extracted from a video taken of the reaction. Images are spaced by approximately 5 seconds. The size of the image is *circa* 24 by 17mm.

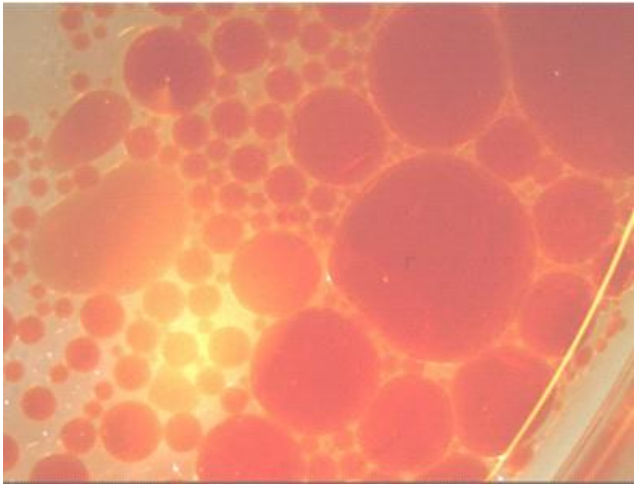

Figure 2a

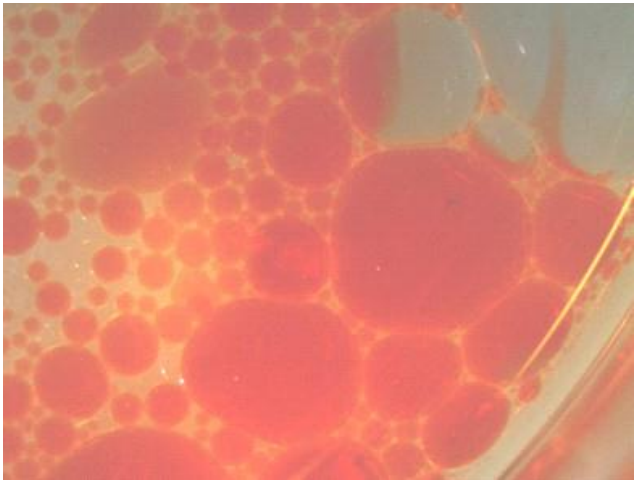

Figure2b

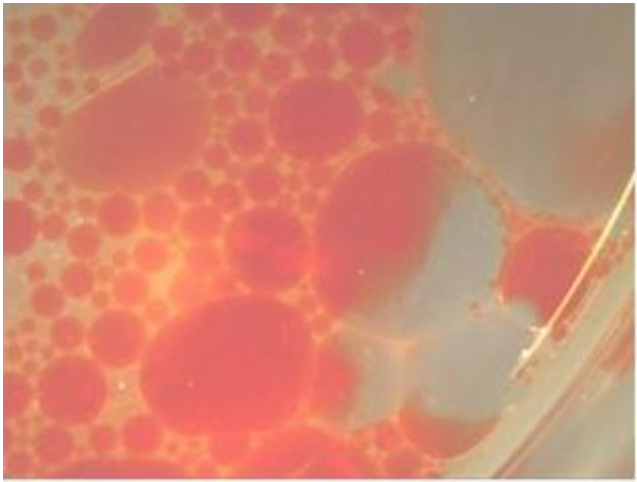

Figure 2c

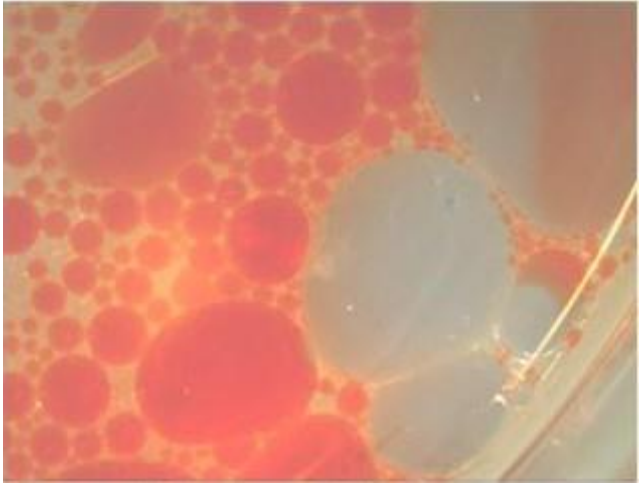

Figure 2d

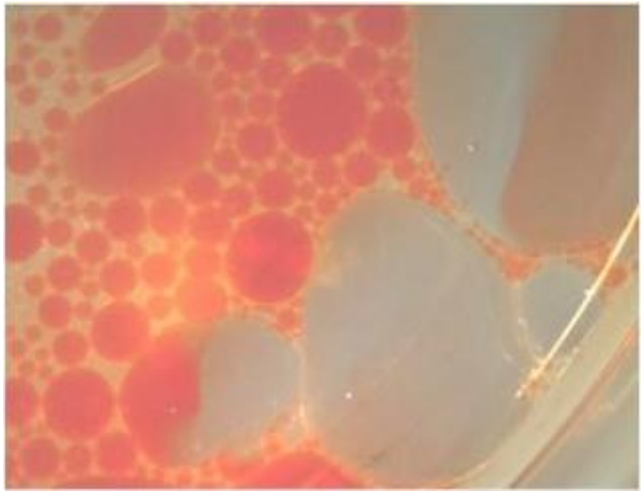

Figure 2e

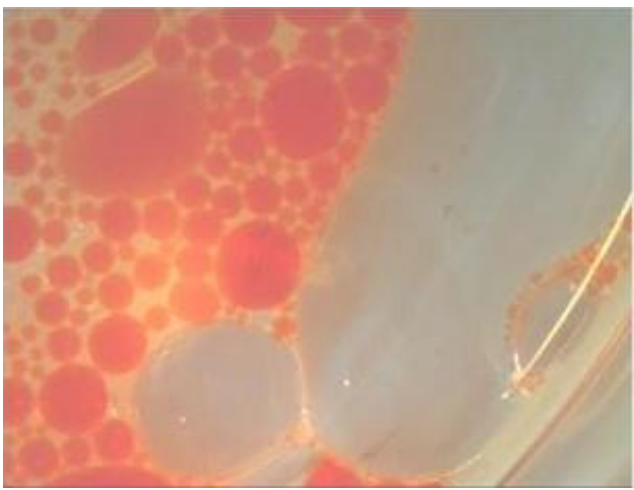

Figure 2f

Figure2. Showing vesicle amalgamation in the wake of oxidation fronts. This reaction had a higher concentration of acid (0.20ml of 3M sulphuric acid, 0.50ml of 1.5M sodium bromate, 0.875ml of

0.3M CHD, 0.15ml 1M potassium bromide, 0.17ml of 0.025M ferroin and 0.587ml of deionised water) than the more stable reaction shown in figure 1. Image size *circa* 24 by 18 mm.

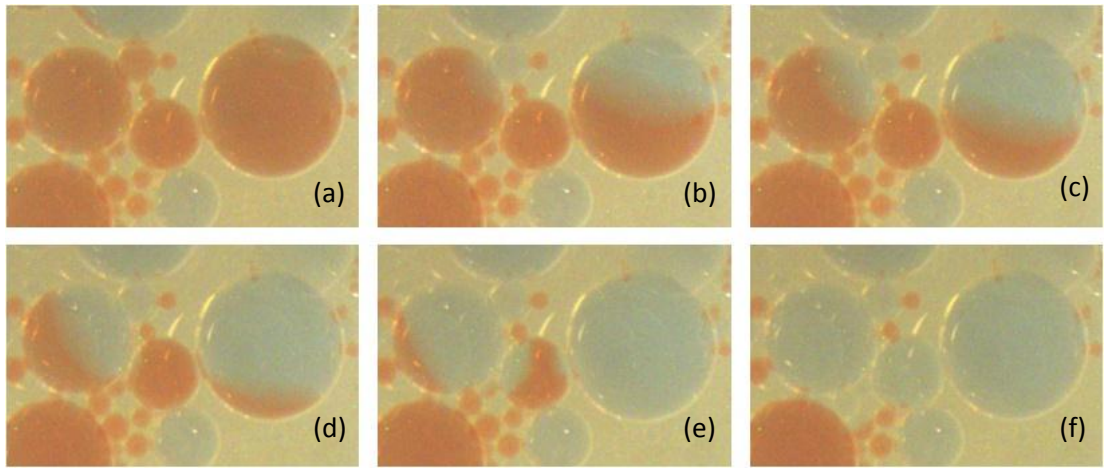

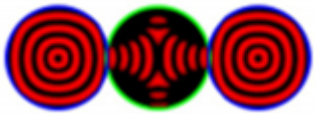

(g)

Figure 3 a-f showing excitation waves transferring from two input vesicles to a central output vesicle. Figure 3g showing a similar scheme implemented in 2D discs (Holley *et al.* 2011a) as a 2 input OR gate (showing case 1,1 > 1). We can presume that if one of the input vesicles was not in the oxidised state that the excitation wave would travel from the oxidised input across the output cell to the other input (giving case 1,0> 1). It would be possible to create an AND gate from a similar arrangement if another vesicle was placed above or below the current output. In this case it would have to be shown that a wave only transfers to the output as a result of wavefront collision. Image size 8 by 5 mm.

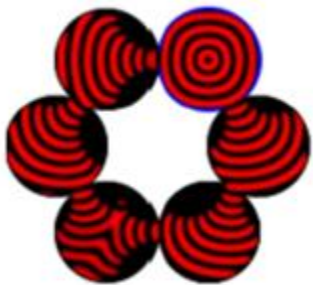 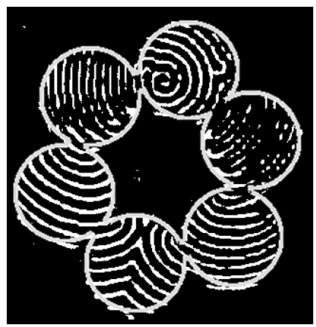

Figure 4a                    Figure 4b

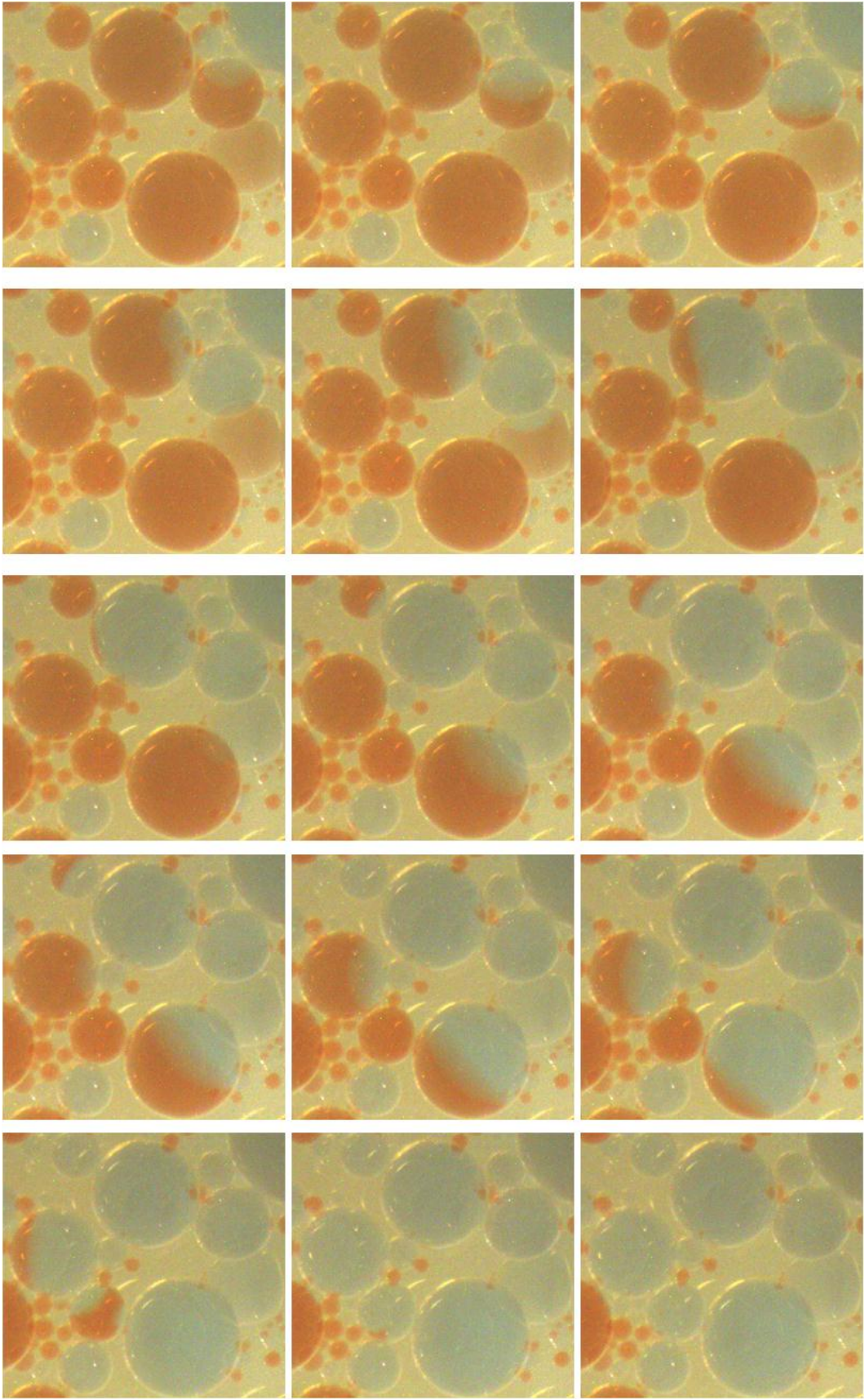

Figure 4c

Figure 4. Towards constructing memory cells in arrays of vesicles. a) Shows the simulation results for the construction of a memory cell from 2D discs (Holley et al. 2011a). b) shows an experimental implementation of the scheme in a 2D light sensitive BZ system (Holley et al. 2011b) c) shows the implementation of a similar scheme in a network of polydisperse vesicles. Image size *circa* 8 by 8mm.

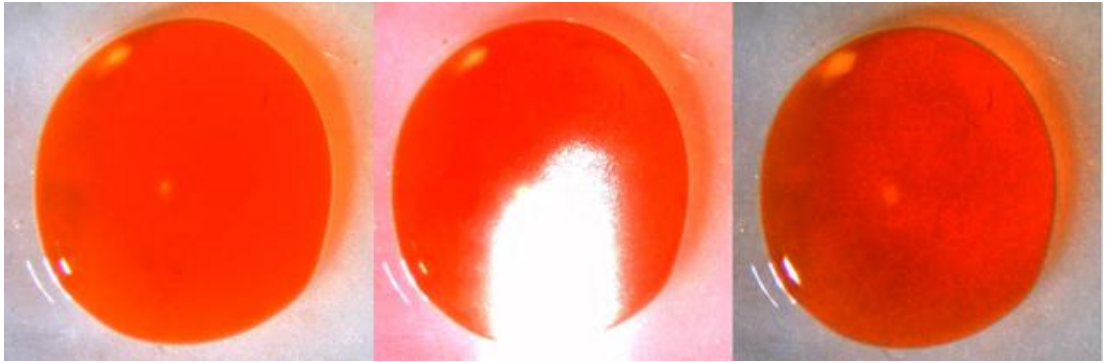

Figure 5. Showing the application of a red laser (650nm) with no evidence of activation. The application of the laser was maintained for 10 seconds. Vesicle diameter *circa* 5mm

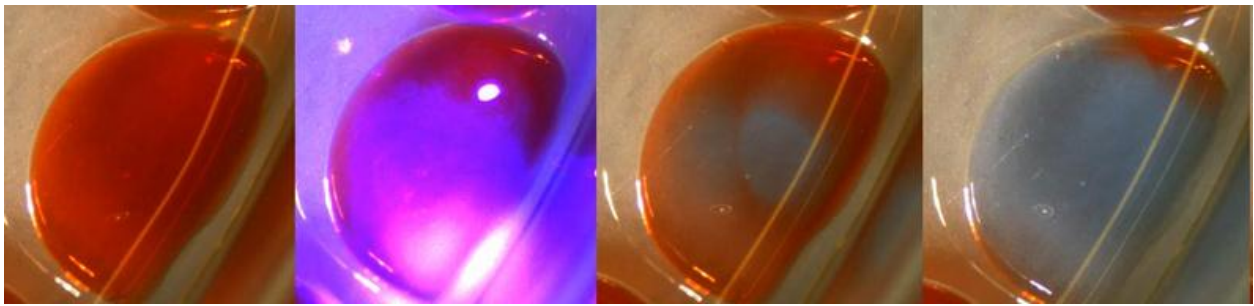

Figure 6. Showing the application of a blue laser (405nm) for 5 seconds with clear evidence of localised activation. Vesicle diameter *circa* 5mm.

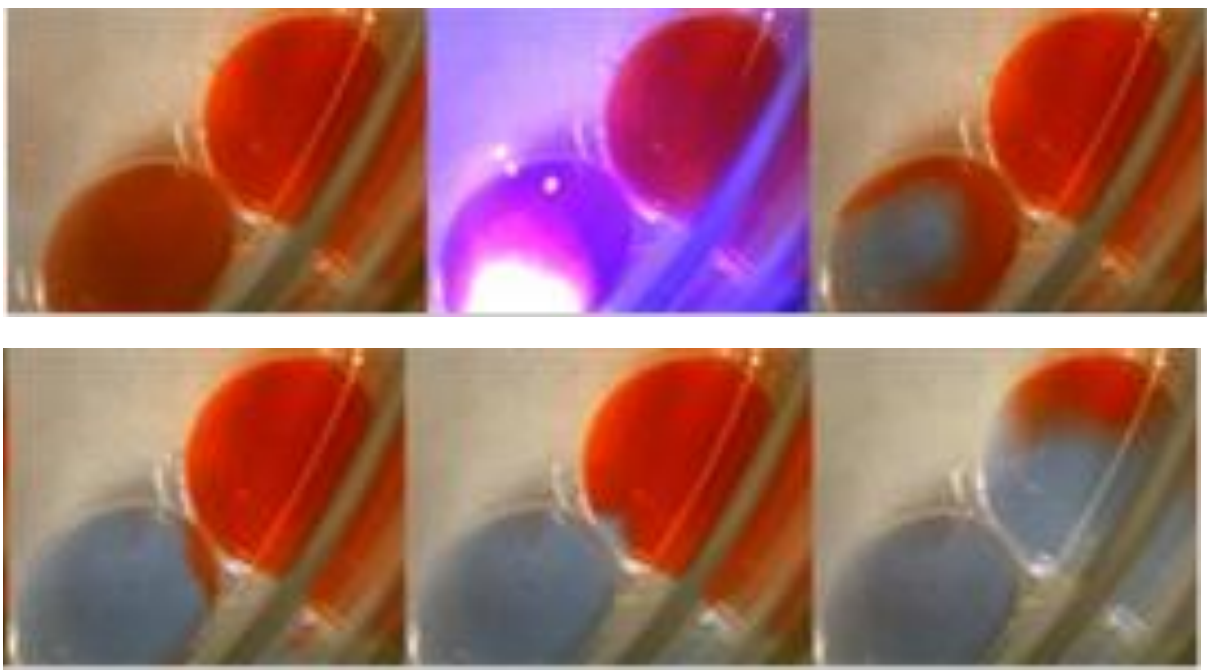

Figure 7. Showing localised initiation of excitation wave in a reduced vesicle via the application of blue laser light and the subsequent transfer of excitation to an adjacent reduced vesicle.

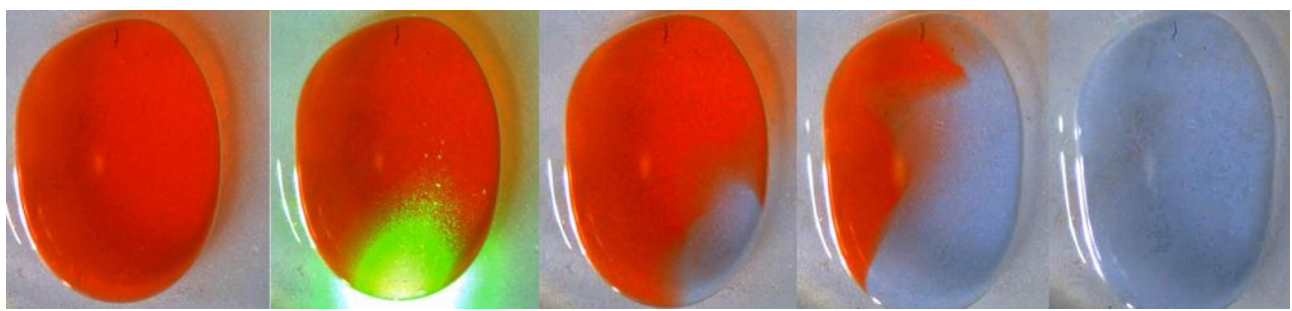

Figure 8. Showing the localised initiation of an excitation wave in a reduced vesicle via the application of a green laser 532nm for 5 seconds. Vesicle diameter *circa* 5mm.

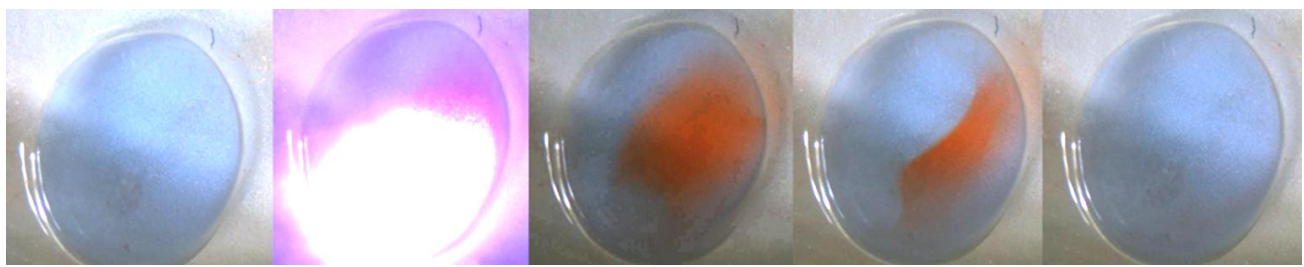

Figure 9. Showing the application of a blue laser to an already oxidised vesicle with clear localised reversion to a reduced state. Vesicle diameter *circa* 5mm.